\documentclass[twoside,11pt]{article}

\usepackage{blindtext}

\usepackage[preprint]{dmlr2e}

\usepackage{booktabs}

\usepackage{xurl}

\usepackage{scalerel}
\usepackage{tikz}
\usetikzlibrary{svg.path}

\usepackage{xcolor}
\usepackage{algorithmic}
\usepackage{graphicx}
\usepackage{textcomp}
\usepackage{pgfplots}
\usepackage{collcell}
\usepackage{hyperref}
\usepackage{microtype}

\usepackage{float}

\usepackage{colortbl}

\usepackage{booktabs}
\usepackage{pbox}
\usepackage{rotating}
\usepackage{tikz}
\usepackage{xparse}
\usepackage{fancyvrb}
\usepackage{moresize}

\usepackage[frozencache,cachedir=.]{minted}
\usepackage{etoolbox}

\usepackage[mode=buildnew]{standalone} 

\makeatletter
\AtBeginEnvironment{minted}{\dontdofcolorbox}
\def\dontdofcolorbox{\renewcommand\fcolorbox[4][]{##4}}
\makeatother

\definecolor{lb}{RGB}{171,219,227} 

\definecolor{col1}{HTML}{FFBE0B}
\definecolor{col2}{HTML}{FB5607}
\definecolor{col3}{HTML}{FF006E}
\definecolor{col4}{HTML}{8338EC}
\definecolor{col5}{HTML}{3A86FF}

\definecolor{col6}{HTML}{EF8534}

\usepackage{todonotes}

\newcommand{\cppLogo}{\raisebox{-1.5pt}{\includegraphics[height=1.05em]{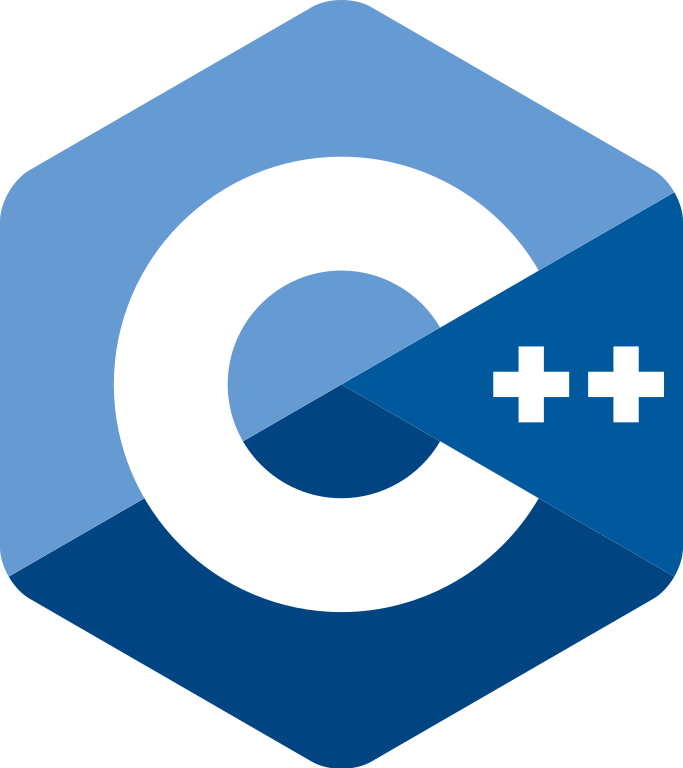}}\xspace}
\newcommand{\cLogo}{\raisebox{-1.5pt}{\includegraphics[height=1.05em]{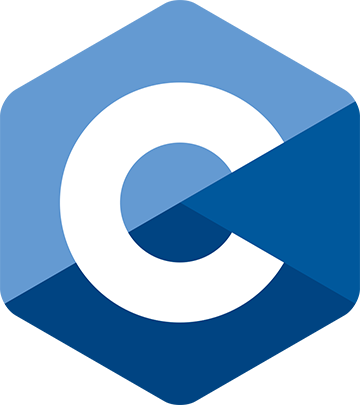}}\xspace}
\newcommand{\rustLogo}{\raisebox{-1.5pt}{\includegraphics[height=1.05em]{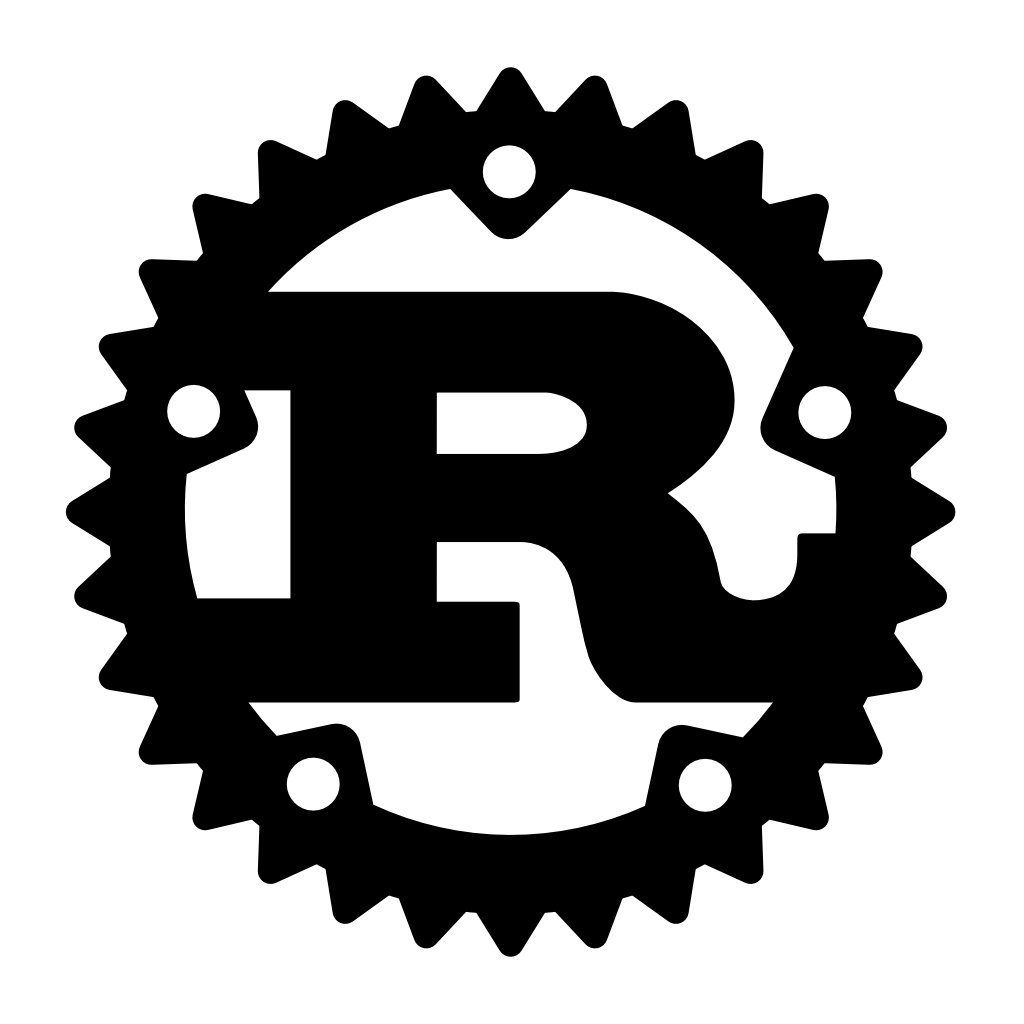}}\xspace}
\newcommand{\juliaLogo}{\raisebox{-1.5pt}{\includegraphics[height=1.05em]{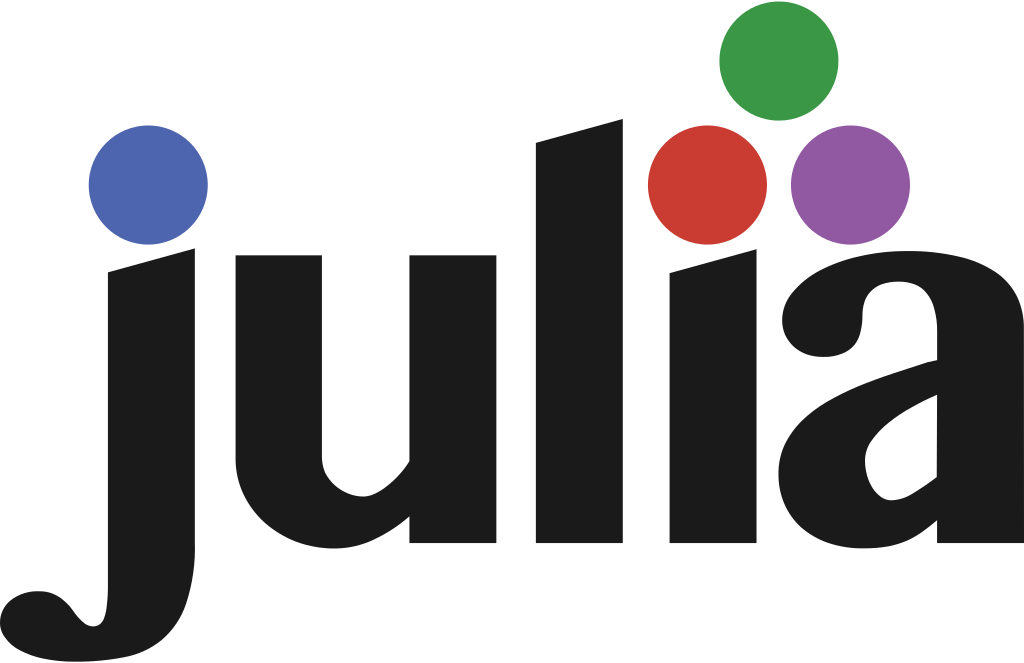}}\xspace}

\newcommand{\swiftLogo}{\raisebox{-1.5pt}{\includegraphics[height=1.05em]{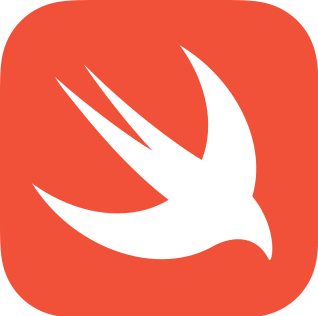}}\xspace}

\usepackage{lastpage}
\dmlrheading{23}{2024}{1-\pageref{LastPage}}{9/23; Revised 10/23}{11/23}{21-0000}{Grossman et al.} 
\def\openreview{\url{https://openreview.net/forum?id=XXXX}} 

\ShortHeadings{ComPile: A Large IR Dataset from Production Sources}{Grossman et al.}
\firstpageno{1}

\begin{document}

\title{ComPile: A Large IR Dataset from Production Sources}

\author{\name Aiden Grossman$^{+,*,1}$ \email amgrossman@ucdavis.edu
       \\
       \name Ludger Paehler$^{*,2}$ \email ludger.paehler@tum.de
       \\
       \name Konstantinos Parasyris$^{3}$ \email parasyris1@llnl.gov
       \\
       \name Tal Ben-Nun$^{3}$ \email talbn@llnl.gov
       \\
       \name Jacob Hegna$^{4}$ \email jacobhegna@gmail.com
       \\
       \name William S. Moses$^{5}$ \email wsmoses@illinois.edu
       \\
       \name Jose M. Monsalve Diaz$^{6}$ \email jmonsalvediaz@anl.gov
       \\
       \name Mircea Trofin$^{7}$ \email mtrofin@google.com
       \\
       \name Johannes Doerfert$^{*,3}$ \email jdoerfert@llnl.gov\\
       ~\\
       \small{$^1$ \textit{University of California, Davis, USA}}\\
       \small{ $^2$ \textit{School of Computation, Information and Technology, Technical University of Munich, GER}}\\
       \small{$^3$ \textit{Center for Applied Scientific Computing, Lawrence Livermore National Laboratory, USA}}\\
       \small{$^4$ \textit{University of Minnesota, Twin Cities, USA}}\\
       \small{$^5$ \textit{Department of Computer Science, University of Illinois Urbana-Champaign, USA}}\\
       \small{$^6$ \textit{Division of Mathematics and Computer Science, Argonne National Laboratory, USA}}\\
       \small{$^7$ \textit{Google Inc., USA}}
}

\editor{My editor}

\maketitle
\def\thefootnote{+}\footnotetext{Work performed while at LLNL}\def\thefootnote{\arabic{footnote}}
\def\thefootnote{*}\footnotetext{Corresponding authors}\def\thefootnote{\arabic{footnote}}
\def\thefootnote{}\footnotetext{Permissibly licensed subset of the dataset available under \href{https://huggingface.co/datasets/llvm-ml/ComPile}{huggingface.co/datasets/llvm-ml/ComPile}}\def\thefootnote{\arabic{footnote}}

\begin{abstract}
Code is increasingly becoming a core data modality of modern machine learning research impacting not only the way we write code
with conversational agents like OpenAI's ChatGPT, Google's Bard, or Anthropic's Claude, the way we translate code from one language
into another, but also the compiler infrastructure underlying the language. While modeling approaches may vary and representations differ,
the targeted tasks often remain the same within the individual classes of models. Yet, relying solely on the ability of modern models to extract
information from unstructured code does not take advantage of 70 years of programming language and compiler development by not utilizing the
structure inherent to programs in the data collection. This detracts from the performance of models working over a tokenized representation
of input code and precludes the use of these models in the compiler itself. To work towards the first intermediate
representation (IR) based models, we fully utilize the LLVM compiler infrastructure, shared by a number of languages, to generate
a $1.4$T Llama 2 token dataset of LLVM IR. We generated this dataset from programming languages built on the shared LLVM
infrastructure, including Rust, Swift, Julia, and C/C++, by hooking into LLVM code generation either through the language's package
manager or the compiler directly to extract the dataset of intermediate representations from production grade programs. 
Statistical analysis proves the utility
of our dataset not only for large language model training, but also for the introspection into the code generation process itself
as well as for training of machine-learned compiler components.
\end{abstract}

\begin{keywords}
  Code, Multilingual, Compiler, LLVM, Intermediate Representation
\end{keywords}

\section{Introduction}
\label{sec:introduction}

With the encapsulation of attention~\citep{chorowski2015attention} in the modern transformer architecture~\citep{vaswani2017attention}, 
the transformer has dominated many natural language processing tasks, starting with the widely used BERT architecture~\citep{devlin2018bert}.
Adjacent fields, such as vision~\citep{alayrac2022flamingo}, and cross-modal models, such as natural language to vision models, all have been transformed
by the modern singular architecture approach.
Originally, due to the immense computational cost of training~\citep{hoffmann2022training}, models with full weights for further training were only sparsely available.
Fine-tuning for downstream tasks~\citep{brown2020language}, or task suites~\citep{srivastava2023beyond}, allows modern large language models to solve a wider array of modeling tasks. In the past few years, there has been a Cambrian explosion
in the availability of pre-trained capable models for fine-tuning. There are currently a large number of open-source release of pre-trained model
series such as OPT~\citep{zhang2022opt}, Llama~\citep{touvron2023llama1}, Llama 2~\citep{touvron2023llama}, Pythia~\citep{biderman2023pythia}, MPT~\cite{portes2023mosaicbert}, and the recently released StarCoder 2~\citep{lozhkov2024starcoder2}.
The wider availability of model weights, architectures, and training checkpoints has enabled the application and tuning of these increasingly capable large language models for domains such as code.

\noindent Beginning with the first BERT models for code~\citep{feng2020codebert} and their extension to graphs~\citep{guo2020graphcodebert},
code has remained a highly active data modality and has seen a constant flurry of new ideas, interfaces, representations, and
downstream tasks. Most recently, the rise of instruction-tuned~\citep{ouyang2022training}, and reinforcement learning-trained
large language models~\citep{christiano2017deep} have enabled a completely new interface to these models. For example, by conversing with a large
language model for code~\citep{chatgpt2023,bard2023}, the user \textit{prompts} the model
with their query, and the model then writes the prompted-for code. This has, in turn, spawned an entire class of new prompting
approaches specifically designed for code, such as grammar-based sampling and sequential Monte-Carlo steering~\citep{lew2023sequential}. Approaches to the
construction of large language models for code vary. Some large language models use a base model, such as, e.g., Code Llama~\citep{roziere2023code},
that is fine-tuned through a general training corpus that only contains code~\citep{openai2023gpt4,claude2023}. On the other hand, other models are only trained on code
from the outset~\citep{copilot2023,roziere2021leveraging,szafraniec2023transcoderir}. We focus on the category of models for which one utilizes a
pre-trained base building block trained on a larger training corpus consisting of not only code, to then
fine-tune on a code-only \textit{code training} corpus. Recently, several models have appeared, such as
Meta's Code Llama~\citep{roziere2023code}, Alphabet's Codex~\citep{codey2023}, WizardCoder~\citep{luo2023wizardcoder}, and the large cross-institutional
collaborations StarCoder~\citep{li2023starcoder,lozhkov2024starcoder2}, and SantaCoder~\citep{allal2023santacoder}. All these share the goal of assisting
people in writing code but miss the opportunities afforded by combining the properties of these powerful modern models (i.e., 
scale, capabilities, and adaptiveness) with the established shared compilation infrastructure that has made programming languages faster, more robust, and easier to use.

\noindent In several pieces of previous work~\citep{ali2020neurovectorizer,kulkarni2013automaticinlining},
this transformative potential was harnessed,
machine-learned heuristic replacements developed, and in some cases~\citep{trofin2021mlgo} the \textit{heuristics were upstreamed to the main LLVM codebase},
improving all code run through LLVM when the ML heuristics are enabled. Orthogonal to the replacement of heuristics
with machine learning, a large number of people have explored the ordering of compiler passes
~\citep{cummins2022compilergym,huang2020autophase}. While the learning of pass orderings was initially held
back by the lack of easy-to-access, high-performance reinforcement learning environments to validate new reinforcement learning
strategies, this has by now been addressed with the introduction of CompilerGym~\citep{cummins2022compilergym}.
In contrast, the learning of entirely new heuristics, optimization passes, and other compiler components with
large language models~\citep{yang2023large,cummins2023llmcompileroptimizations} to realize the transformative potential
of the model class is held back partially by the lack
of large datasets of high-quality code to train such models properly. Models are only trained on smaller datasets, such as
Anghabench~\citep{dasilva2021anghabench}, Exebench~\citep{armengol2022exebench}, and HPCORPUS~\citep{kadosh2023quantifying},
or sometimes rely on synthetic benchmarks. Synthetic benchmarks can be aided by ML techniques~\citep{cummins2017synthesizingbenchmarks,cummins2023benchdirect}
and even closely match some properties of the corpora they are trained on, but these techniques themselves still suffer from a small training
set and can only approximate the properties of production code. Small datasets lead to smaller, worse-performing
models and hence do not allow such compiler-focussed models to fully access the fine-tuning paradigm utilized by 
modern large language models with their base models of multiple billion parameters. 

\subsection{Contributions}

\begin{figure}[t]
\centering
\includegraphics[width=0.6\textwidth]{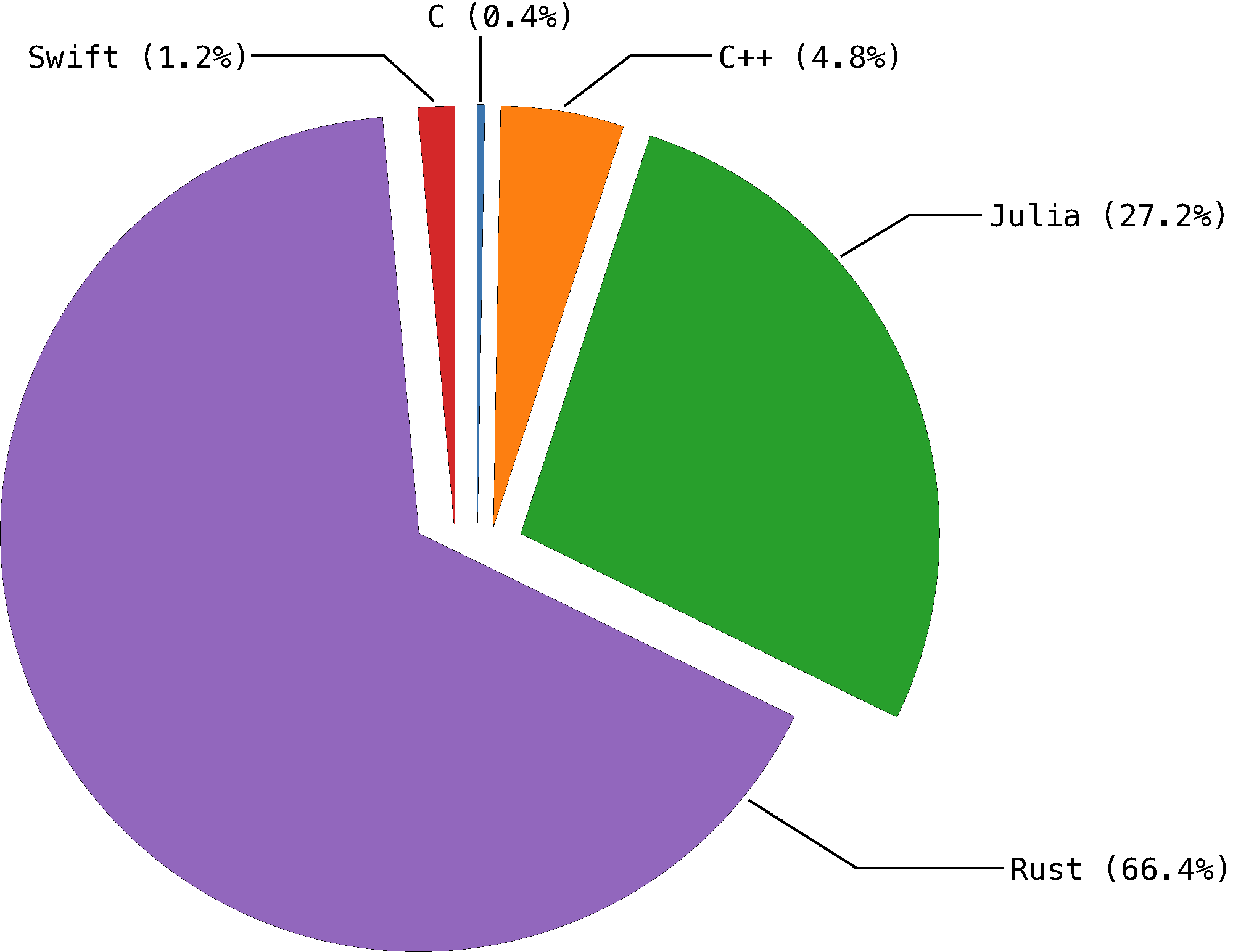}
\caption{
    Size distribution of LLVM intermediate representation (IR) bitcode within ComPile before de-duplication within and among languages.
}
\label{fig:size_treemap}
\end{figure}

Focusing on the paradigm
of taking a pre-trained basic building block, we pose the question \textit{"What does a modern, large code training dataset for
compilers actually look like?"} and construct a high-quality dataset of a similar scale to existing LLM datasets solely at the level of compiler
intermediate representation. Within this context, we associate quality with the \textit{usage} of code, with code being
used more often being of higher quality for our purposes. Correctly being able to reason about very widespread code in
production systems is incredibly important for compiler work. 
In the short term, we believe our dataset will enable the training of larger language models for
compilers useful for a broader array of downstream tasks after fine-tuning, and in the long-term enable use-cases such as direct performance
prediction to obtain a reliable runtime estimate without ever needing to run a single line of code. To these goals, our paper makes the following contributions:
\begin{itemize}
    \item The {\bf{introduction of a 2.8TB dataset}} of textual intermediate representation of the shared LLVM compiler infrastructure encompassing {\bf{production-grade programs from Rust, Julia, Swift, and C/C++}}. A broad overview of the size distribution is shown in \autoref{fig:size_treemap}.
    \item Preliminary {\bf{large scale statistical analyses of LLVM-IR modules across multiple languages}}, demonstrating
    the utility of our dataset and tooling.
    \item Demonstration of the utility of ComPile for training large machine learning models through {\bf{quantification of size
    and approximate token counts}}.
    \item {\bf{Open-sourcing of workflow and compiler tooling}} to construct massive-scale code datasets, which are easy to install and ready for scalable deployment in HPC and cloud environments. The statistics of the entire dataset constructable with the tooling are available in the appendix~\ref{appendix:closet_dataset_stats}.
\end{itemize}

\begin{table}[h!]
    \centering
    \begin{tabular}{ccccccc}
        \multicolumn{7}{c}{\bf{\large{Public Dataset}}} \\
        \toprule
        & \multicolumn{2}{c}{} & \multicolumn{4}{c}{\bf{BPE Tokens}} \\
        \cmidrule(l{5pt}r{5pt}){4-7}
         \multicolumn{2}{c}{\bf{Language}} & {\renewcommand{\arraystretch}{1}\begin{tabular}[c]{@{}c@{}}\textbf{\textbf{Llama 2 Tokens}}\\\textit{(billions)}\vspace{.0em}\end{tabular}} & {\renewcommand{\arraystretch}{1}\begin{tabular}[c]{@{}c@{}}\textbf{\textbf{10k}}\\\textit{(billions)}\vspace{.0em}\end{tabular}} & {\renewcommand{\arraystretch}{1}\begin{tabular}[c]{@{}c@{}}\textbf{\textbf{50k}}\\\textit{(billions)}\vspace{.0em}\end{tabular}} & {\renewcommand{\arraystretch}{1}\begin{tabular}[c]{@{}c@{}}\textbf{\textbf{50k}}\\\textit{(billions)}\vspace{.0em}\end{tabular}} & {\renewcommand{\arraystretch}{1}\begin{tabular}[c]{@{}c@{}}\textbf{\textbf{100k}}\\\textit{(billions)}\vspace{.0em}\end{tabular}} \\
         \midrule
         \cLogo & C & 5 & 1 & 1 & 0.5 & 0.4 \\
         \cppLogo & C++ & 47 & 11 & 6 & 5 & 4 \\
         \juliaLogo & Julia & 548 & 42 & 23 & 18 & 12 \\
         \rustLogo & Rust & 736 & 137 & 90 & 79 & 69 \\
         \swiftLogo & Swift & 20 & 3 & 2 & 1 & 1 \\
         \midrule
         \multicolumn{2}{c}{\textbf{Total}} & \textbf{1355} & \textbf{195} & \textbf{122} & \textbf{104} & \textbf{88} \\
         \bottomrule
    \end{tabular}
    \caption{
        Token count of the encoded ComPile under varying vocabulary sizes, and considering the tokenization of the data with Byte-Pair encodings \emph{(BPE)}, and tokenization with the Llama 2 tokenizer.
    }
    \label{stats:public_tokens}
\end{table}

\section{Background}
\label{sec:background}

Building upon package ecosystems as sources of intermediate representation is ideal due to the large amount of packaged high-quality code and the abstraction over the build systems of individual projects. This abstraction is due to a common build wrapper that
invokes the individual build systems with the relevant configuration options. Package managers
are designed to install a set of packages that a user desires, abiding by some constraints
from some repositories. Each package manager often has its own repositories that
are built from source. The recipes used to build the included applications often specify exact build steps to build a piece
of software, including an exact and consistent specification of dependencies needed to build said software. These
package recipes can also often be modified to perform some additional steps or to modify the build process
itself. Some build systems, such as cargo, are combined with package managers, allowing them to build a piece
of software and all of its dependencies that the build system supports installing itself. Modifying these build processes
allows us to take advantage of the dependency management and other aspects of build recipes already present for a
significant number of packages. However, many build systems do not explicitly support custom modification of build
recipes or build-time configuration options, including compile flags. In this work, we choose to specifically focus
on utilizing package managers that explicitly allow setting compiler flags, such as the from-source package manager Spack \citep{gamblin2015spack} that is focused on high-performance computing (HPC).

\begin{table}[t]
\centering
\begin{tabular}{ccccc}
\toprule
{\bf{Language}} & {\bf{Source Code}} & {\bf{Unoptimized IR}} & {\bf{Optimized IR}} & {\bf{X86 Assembly}} \\
\midrule
C &
\begin{minipage}{2.6cm}
\begin{minted}[fontsize=\ssmall]{c}
int sum(int a, int b)
{
    return a+b;
}
\end{minted}
\end{minipage} &
\begin{minipage}{2.8cm}
\vspace{2pt}
\begin{minted}[fontsize=\ssmall]{llvm}
define i32 @sum
(i32 %0, i32 %1) {
  %3 = alloca i32
  %4 = alloca i32
  store i32 %0, ptr %3
  store i32 %1, ptr %4
  %5 = load i32, ptr %3
  %6 = load i32, ptr %4
  %7 = add i32 %5, %6
  ret i32 %7
}
\end{minted}
\end{minipage} &
\begin{minipage}{2.8cm}
\begin{minted}[fontsize=\ssmall]{llvm}
define i32 @sum
(i32 %0, i32 %1) {
  %3 = add nsw i32 %1, %0
  ret i32 %3
}
\end{minted}
\end{minipage} &
\begin{minipage}{2cm}
\begin{minted}[fontsize=\ssmall]{gas}
sum:
  push    rbp
  mov     rbp, rsp
  mov     eax, esi
  add     eax, edi
  pop     rbp
  ret
\end{minted}
\end{minipage}
\\
\midrule
Rust &
\begin{minipage}{2.6cm}
\begin{minted}[fontsize=\ssmall]{rust}
pub fn sum(
    a: i32, b: i32
) -> i32 {
  a + b
}
\end{minted}
\end{minipage} &
\begin{minipage}{2.8cm}
\vspace{2pt}
\begin{minted}[fontsize=\ssmall]{llvm}
define i32 @a::sum
(i32 %a, i32 %b) {
start:
  %_0 = add i32 %a, %b
  ret i32 %_0
}
 \end{minted}
 \end{minipage} &
 \begin{minipage}{2.8cm}
 \begin{minted}[fontsize=\ssmall]{llvm}
define i32 @a::sum
(i32 %a, i32 %b) {
start:
  %_0 = add i32 %a, %b
  ret i32 %_0
}
 \end{minted}
 \end{minipage} &
 \begin{minipage}{2cm}
 \begin{minted}[fontsize=\ssmall]{gas}
example::sum:
  mov     eax, edi
  add     eax, esi
  ret
 \end{minted}
 \end{minipage} \\
 \bottomrule
\end{tabular}
\caption{
    The transformations source code goes through into assembly through the compiler's LLVM intermediate representation. We collect the intermediate representation at the unoptimized stage.
}
\label{table:example_compilation}
\end{table}

~\\
\noindent In addition to utilizing package managers, we also take advantage of several aspects of the LLVM compilation infrastructure \citep{lattner2004llvm},
particularly the Clang C/C++ frontend and LLVM-IR, the intermediate representation LLVM uses.
The full process of compilation, such as the one performed by Clang with LLVM during the compilation of C/C++, is
composed of three main stages: the frontend, the middle-end, and the backend. The entire compilation process is exemplified
in \autoref{table:example_compilation}. A compiler frontend has the job
of taking a piece of source code, typically a single source file, sometimes called a translation unit, and 
generating a \textit{module} of intermediate representation that can then be processed by a compiler middle-end,
such as LLVM. A module typically contains multiple functions, referenced globals, and relevant metadata.
Compiler intermediate representations, or IRs, are designed to sit between the source programming language and the compiler's output, assembly. They are typically designed to be source-language and target-agnostic. This allows code written to 
modify and process the IR to be reused across many languages and target platforms. IRs typically also have additional properties that make
them particularly amenable for performing optimizations. LLVM-IR specifically enforces single static assignment (SSA), where all
variables are assigned exactly once and referenced multiple times. This makes certain analyses much easier, such as dataflow analysis.
LLVM uses its intermediate representation, LLVM-IR, to perform optimizations and other operations related to
lowering source code to machine code in a manner that abstracts away most details of the target machine. Within LLVM,
the compiler middle-end operates over the IR produced by the frontend through a series of grouped operations called passes. A \textit{pass} is designed to
perform a specific task, such as removing dead code, simplifying the control flow graph, or combining instructions that
can be simplified. A \textit{pass pipeline} is typically language and optimization-level specific. It comprises a set of passes in a specific
order run over the IR to optimize it for the desired properties.
After optimization, the compiler backend takes over, performing the necessary tasks to transform the (mostly) target-agnostic IR
into target-specific machine code that can be executed on the target machine. The backend typically performs tasks such as instruction selection, instruction
scheduling, and register allocation. In addition, compiler backends also often perform some small target-specific optimizations, such as peephole optimizations, to further improve the characteristics of generated code.
We compose our dataset, \textit{ComPile}, of LLVM-IR, as it gives a common framework across programming languages and target platforms
while also allowing us to perform a detailed analysis of the compiler middle-end. These properties and more make LLVM-IR
a great modality for a compiler-centric dataset useful for compiler tasks such as program analysis, optimizations, and code generation.
\section{Dataset Construction}
\label{sec:dataset_constructions}

\begin{figure}[t]
\centering
\includegraphics[width=\textwidth]{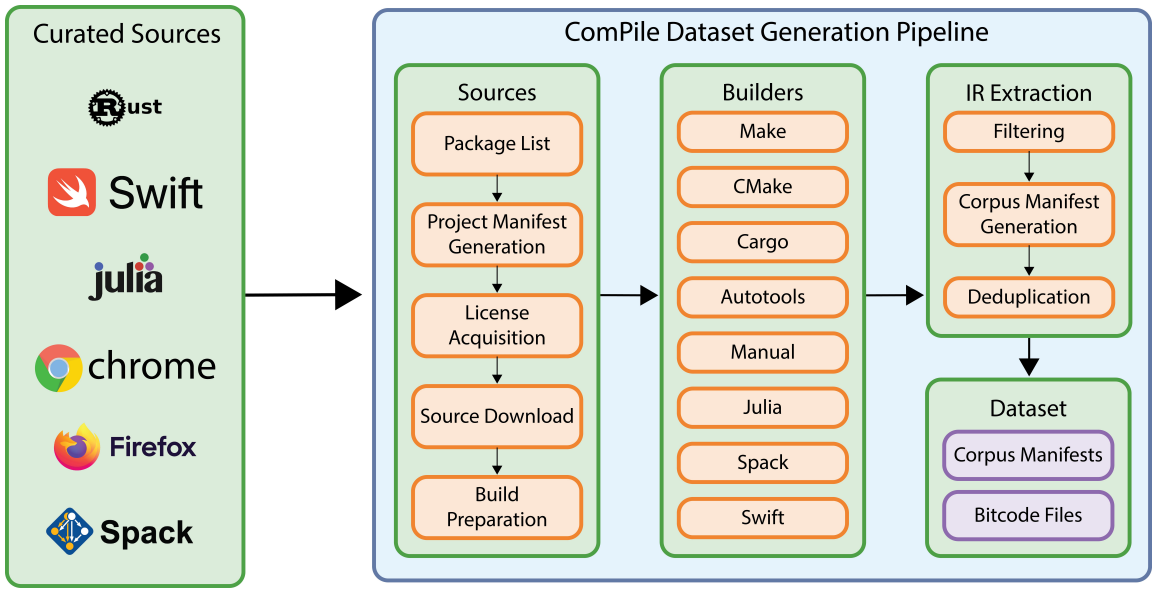}
\caption{
    Individual components of the dataset collection tooling. \emph{(Curated Sources)} The set of sources comprised of package indices, and selected packages, ingested by the ComPile Dataset Generation Pipeline. \emph{(Sources)} acquire the source based upon the provided package list, before the \emph{(Builders)} built the package, and it is then filtered, deduplicated, and its build process documented in the \emph{(IR Extraction)} to arrive at the dataset.
 }
\label{fig:pipeline}
\end{figure}

The building of entire language ecosystems introduces its very own kind of problems, such as ``How
can we build all packages while only manipulating a single builder file?'', ``How can we distribute
the build process of all these packages across many nodes?'' and in the case of just in time (JIT)
compiled languages ``How well defined is the compilation of entire packages?''. In this section, we 
describe in detail our workflow and all modifications to the build systems of individual
languages. A summary of our workflow can be seen in \autoref{fig:pipeline}. To construct the IR
database, we use a set of curated sources focusing on code used in production systems. Individual sources are
defined in \texttt{.json} files. While most projects are hosted in repositories on
GitHub, we also added sources consisting of archived compressed source codes such as tarball files.
The builders then ingest the information from the project on its build system, either through
the manifest information, which contains the information on the building mechanism and commands,
or through an ecosystem specific manifest processed by a script into a complete package manifest.
Next in the workflow is the LLVM-IR extraction. Extracting IR depends on the way the IR is presented in the curated source,
as we will be described below. A deduplication stage removes exact or near duplicate IR to
eliminate common IR, or IR that does not improve the quality of our corpus, which has established
precedent in literature~\citep{allamanis2019adverse}. A manifest that contains a list of LLVM bitcode
modules extracted from the project is then created. Leaning into the shared LLVM compiler infrastructure, we are able to
take advantage of existing LLVM tools and LLVM passes to obtain information about the LLVM-IR modules.
After building, IR extraction, and deduplication, the dataset is then ready for downstream usage in analysis 
or training capacities.~\footnote{Scripts and builders to reproduce the entire dataset are available under:
\href{https://github.com/llvm-ml/llvm-ir-dataset-utils}{github.com/llvm-ml/llvm-ir-dataset-utils}}

\subsection{Ecosystem-Specific Builders}
\label{subsec:ir_generation}

We support extracting LLVM-IR from several large package ecosystems through many different
builders that each handle a specific ecosystem or build system. The major builders that have been implemented
are described below.

\subsubsection{Rust}
For extracting IR from Rust packages, we first extract a list of crates from the Rust Package Repository~\citep{rust2023packagerepo}.
The Rust Package repository has over 100,000 Rust packages that are all buildable using a consistent build system, \texttt{cargo},
which makes the build process very feasible to automate. Additionally, the process of listing packages in the
repository serves as a filter, preventing the ingestion of unused or experimental Rust code,
which could otherwise feasibly impact the IR distributions within the final corpus.
We prioritize building crates from the indicated git repository in the package repository as we found that git
repositories often have additional targets not included in the tar archives uploaded to the package repository
that yield more IR. We remove crates that point to the same git repository to improve build times and prevent
excessive duplication within the dataset. In addition to pulling from a git repository, we also use the package
repository provided tar archive as a fallback as we found that many git repositories would fail to clone. For
building each crate, we used the native cargo build system. We first extract a list of targets and sub-packages
from the package manifest and then built each one with \mintinline{Bash}{cargo rustc}, specifying the appropriate
sub-package and target and passing the \mintinline{Bash}{--emit=llvm-bc} flag to make cargo additionally generate
LLVM IR. We built Rust packages without any optimizations to ensure we get unoptimized bitcode. This could feasibly
impact the distribution of our dataset as Rust has two high level Rust-specific IRs that are used to optimize Rust
code before it is lowered into LLVM-IR that do not perform optimizations without optimization flags. We leave analysis
of distribution shifts related to pre-LLVM optimizations to future work.

\subsubsection{Julia}
To extract code from Julia, we used the official package registry \citep{JuliaRegistriesRepo2023} as a source of
over 9000 packages. We then processed them using a custom pipeline to extract bitcode. Due to the nature of Julia's
``ahead-of-time just-in-time'' (AOT-JIT) design, the ``compilation'' of a whole package is an ill-defined task. Julia only completely lowers a
function to IR at runtime when the function is called. This makes the extraction of bitcode for an entire package
difficult as every single function within a package has to be run with all potential function signatures. However,
there has recently been a large push to precompile packages using \texttt{PackageCompiler.jl}~\citep{PackageCompilerDocs2023}.
The precompilation process involves completely compiling a variety of often used functions within a package and
caching them for later use so the user experiences less wait time when using the package later. This is performed
automatically for many packages during the installation process. In addition to taking advantage of IR generated
during the precompilation process, we also run unit tests, if available, to force the lowering of additional functions.
To grab bitcode per package, we implemented a custom hook in the compilation process\footnote{\href{https://github.com/JuliaLang/julia/pull/50946}{github.com/JuliaLang/julia/pull/50946}}.
We subsequently post-process the produced files to only grab IR which contains actual code rather than serialized
Julia data structures. All the IR files for a package are gathered with the dependencies to capture all uses
of a function. This results in duplicate code, which is removed through a deduplication pipeline. This process inevitably
leaves some gaps in the collected IR. For example, we are not capturing the function specializations exactly used in
production code, instead capturing the function specializations deemed important by the package authors in the
package compilation step and from the available unit tests. However, we believe these nuances of the collection
process do not impact the results presented in this paper or the utility of the collected Julia code significantly.

\subsubsection{Swift}
To prepare a list of Swift packages, we used the Swift Package Index \citep{SwiftPackageIndex2023}, processing all
the GitHub repositories present in the packages list. We then cloned and built every repository we found
with Swift. Swift automatically resolved the dependencies that it was able to and, by passing the flags
\mintinline[breaklines]{bash}{--emit-swift-module-separately -Xswiftc --embed-bitcode}, we are able to embed bitcode within
the object files produced during the compilation process for extraction later. To embed the bitcode correctly,
we have to use the \mintinline{bash}{--emit-swift-module-separately} flag to deactivate the default behaviour
of \mintinline{bash}{swiftc} to emit partial modules, and only merging them later, which is incompatible with
bitcode embedding. While we were able to get some packages to build,
our tooling is designed for a Linux environment, and not the Swift-preferred MacOS environment. Although Linux has
Swift platform support, it does not have support for several of the closed-source dependencies that many Swift
packages require such as SwiftUI.

\subsubsection{Spack}
To include HPC packages, we utilized the HPC package manager Spack~\citep{gamblin2015spack}.
Spack contains a large set of packaged applications, many of them C/C++, and it lets the user specify
the compiler toolchain and any compiler flags to use. In addition, the packaging process in Spack
serves as a quality filter for the dataset, as Spack selects for only those HPC packages whose developers
or users have opted to take steps to contribute their software to Spack. Getting a package into Spack requires
review on GitHub, and this tends to select for popular HPC packages that people {\it want} to use. While Spack
also contains packages that use a significant amount of Fortran, we only extract bitcode generated from C/C++,
because most Fortran packages are not yet compatible with recent LLVM-based Fortran frontend.

\noindent To build a corpus of IR from Spack, we start by extracting a list of packages. Spack supports a variety of
different package types, including Python packages and custom build systems, many of which will not produce IR.
We filter the packages by build system, including only packages that use the common C/C++ build systems CMake, Meson,
Autotools, and Makefiles. After we have a list of packages, we then \textit{concretize} each package. {\it Concretization}
is the process of generating the fully satisfied dependency graph (including flags, build options, microarchitecture targets,
and optional dependencies) for each package. Spack can optionally unify the dependency graph for packages to ensure
that each dependency is built only once, in one configuration, but we choose to concretize each package separately to
allow each package to have its own dependency configuration. This aids in error handling, and it allows us to run the
concretization process in parallel. If any package is incompatible with another for a {\it unified} set of packages,
concretization fails. If we allow packages to have their own dependency versions, we can split up the process and handle
individual failures more gracefully. However, this methodology leaves us with many duplicate packages in addition to
dependencies that won’t produce any bitcode. We handle this by building all packages in the same manner and passing all
extracted bitcode through a deduplication pipeline.

\noindent After concretization, we build all packages with a custom build distribution system, starting with leaf dependencies and continuing on as more and more packages have all of their dependencies built. We build each package with \mintinline{Bash}{clang} while passing the compiler flags \mintinline{Bash}{-Xclang -fembed-bitcode=all}, which causes LLVM IR to be embedded as bitcode within the generated object files. To extract IR from built packages, we direct Spack to keep the build directory (which contains .o files, libraries, and other build artifacts) by passing the \mintinline{Bash}{--keep-stage} flag. To allow for multi-node parallelism, we take advantage of Spack’s buildcache feature, pushing all built packages to a buildcache so that any node within an allocation can use a built package as a dependency. This allows us to distribute builds across a large cluster and obtain a high degree of parallelism, significantly reducing overall build time for the corpus.

\subsubsection{Individual Packages}
In addition to collecting a significant number of packages available through specific ecosystems, we also
wrote additional tooling to allow for the collection of bitcode from individual curated packages. We wrote
scripts to build applications that use CMake, Autotools, and any other build system that can be invoked
through raw shell commands. These scripts work by invoking the build system using the user-provided arguments
along with some additional flags. These additional flags include setting the compiler to \mintinline{bash}{clang}
to make bitcode extraction possible and passing \mintinline{bash}{-Xclang -fembed-bitcode=all} as C/C++ flags
to ensure that bitcode was inserted into the generated object files. The bitcode is then extracted from the object
files after the build completes which is available for further analysis. We collected bitcode from several large
applications not included in the existing package ecosystems that we deemed to be high impact including 
Chromium, Firefox, and the Linux Kernel. These programs each consist of upwards of tens of gigabytes of
bitcode and contain production code that is run virtually everywhere.

\subsection{LLVM-IR extraction}
The aim of our IR extraction approach is to extract IR immediately after the frontend, before any LLVM optimization
passes have run. This allows us to perform analysis on the IR emitted directly after the frontend, and anywhere
else in the optimization pipeline, as we can perform optimization manually, and introspect the optimization pipeline itself.
It is important
to note that there are certain languages, like Rust and Julia, that use language-specific higher level intermediate
representations for optimizations and other transformations specific to the specific language that we are not able
to introspect with this approach. The process for extracting IR directly after the frontend differs significantly
depending upon the language with the necessary options and configurations for doing so being reported in the build
processes above. After the build process completes for a specific package we are left with an assortment of bitcode
in two different formats depending upon the build system: bitcode embedded in object files or a collection of separate
bitcode files. To extract the bitcode into a structured corpus, we take advantage of the \texttt{ml-compiler-opt} tooling from
MLGO~\citep{trofin2021mlgo} as it is production-proven, and allows for the extraction of IR object files by
analyzing a structured compilation command database, or alternatively by searching for all object files within the
build directory. In addition, it also supports creating a structured corpus from raw bitcode files by searching
the build directory. The exact strategy used is dependent upon the build system. Julia, and Rust directly
emit bitcode. Spack, CMake, Autotools, and manual builds are all currently set up to embed bitcode in object files,
but only CMake is able to provide a structured database of compilation commands. Swift embeds bitcode but needs
additional flags during IR extraction due to the bitcode section naming within the object file differing from clang's.
During IR-extraction we do not strip debug information if it is present as it can easily be stripped later and some
models need to have some debug information in their training corpus to be robust against it. Some builders emit
debug information more commonly than others, such as Rust where we compile in debug mode by default to disable optimizations,
but ultimately whether or not debug information is present is project dependent. Finally, we specifically collect bitcode
rather than textual IR as LLVM supports reading bitcode produced by older versions of LLVM but has no such support
for textual IR, which is also easily produced by running \mintinline{shell}{llvm-dis} over the collected corpus.

\begin{table}[t]
\centering
\begin{tabular}{c@{\hskip .3em}lcccc}
 \toprule
 \multicolumn{2}{c}{\bf{Programming Language}} & {\renewcommand{\arraystretch}{1}\begin{tabular}[c]{@{}c@{}}\textbf{\textbf{Bitcode}}\\\textit{(GB)}\vspace{.0em}\end{tabular}} & {\renewcommand{\arraystretch}{1}\begin{tabular}[c]{@{}c@{}}\textbf{\textbf{Deduplicated}}\\\textbf{\textbf{ Bitcode}}\\\textit{(GB)}\vspace{.0em}\end{tabular}} & {\renewcommand{\arraystretch}{1}\begin{tabular}[c]{@{}c@{}}\textbf{\textbf{Licensed}}\\\textbf{\textbf{ Bitcode}}\\\textit{(GB)}\vspace{.0em}\end{tabular}} & {\renewcommand{\arraystretch}{1}\begin{tabular}[c]{@{}c@{}}\textbf{\textbf{Licensed}}\\\textbf{\textbf{ Text}}\\\textit{(GB)}\vspace{.0em}\end{tabular}} \\
 \midrule
 \cLogo & C & 16 & 8 & 2 & 10 \\
 \cppLogo & C++ & 109 & 74 & 29 & 103 \\
 \juliaLogo & Julia & 200 & 184 & 164 & 1088 \\
 \rustLogo & Rust & 656 & 580 & 400 & 1524 \\
 \swiftLogo & Swift & 8 & 7 & 7 & 36 \\
 \midrule
 \multicolumn{2}{c}{\textbf{Total}} & \textbf{990} & \textbf{853} & \textbf{602} & \textbf{2761} \\ 
 \bottomrule
\end{tabular}
\caption{
    Amount of bitcode contained in the public version of ComPile before and after deuplication, and the size of the bitcode and
    associated textual IR for the public version of ComPile.
}
\label{construction:dataset_size}
\vspace{-0.45cm}
\end{table}

\subsection{Deduplication}
Training dataset deduplication can be important for the performance of several key model characteristics
\citep{allamanis2019adverse, kandpal2022deduplicating}. To this end, we deduplicate the entire dataset 
presented in this paper at the module level by computing a combined hash of all
global variables and functions. To perform the hashing, we upstreamed the
\mintinline{c++}{StructuralHash} to LLVM through the \mintinline{c++}{StructuralHashPrinterPass}
\footnote{\href{https://reviews.llvm.org/D158217}{reviews.llvm.org/D158217}} 
\footnote{\href{https://reviews.llvm.org/D158250}{reviews.llvm.org/D158250}} \footnote{\href{https://reviews.llvm.org/D158317}{reviews.llvm.org/D158317}}.
The structural hashing process only captures semantic details of the IR making it invariant against all changes
that do not impact the meaning of the IR other than function call names. In addition, the implementation does not
capture all semantic information, currently ignoring details such as attributes and instruction dependencies which ensures that near-duplicates may be matched as well. We chose to deduplicate at the module level as this ensures the
majority of the duplicate code is removed from the dataset while leaving all significant context within each
module for performing module-level tasks. This deduplication strategy prevents some tasks from being performed,
such as project-level tasks, which rely on a complete set of modules or metadata. The amount of data removed from the dataset through the deduplication process is heavily language
dependent. For example, Julia, a language where our decision to include bitcode from package dependencies significantly increases the
duplication rate, has a duplication rate of approximately 40\%. Other languages have significantly lower
duplication rates.

\subsection{Dataset Size}

To analyze the size of the dataset, we directly gather the size of all bitcode files in the corpus
before and after deduplication. A 35\% reduction in dataset size after
dedupliction is observed.
While measuring the size of bitcode files gives some idea of the total size, it does not allow for proper
size comparisons to other datasets as LLVM bitcode is highly compressed. To this end, we also compute
the size of all textual IR in the dataset by measuring the size of all disassembled bitcode files. We find
that the size of textual IR is approximately 4.6 times the size of the equivalent bitcode. Precise size
figures for ComPile are available in \autoref{construction:dataset_size}.

\subsection{License Filtering}
\label{subsec:license_filtering}

To filter our closed-source dataset for permissively licensed projects, we filter the entire database of projects compiler into ComPile for the \texttt{MIT}, \texttt{Apache-2.0}, the \texttt{BSD-3-Clause}, and the \texttt{BSD-2-Clause} licenses. For this we obtain the license information from package repositories, GitHub, and in part manually using the \texttt{go-license-detector}~\footnote{\url{https://github.com/go-enry/go-license-detector}}, and distribute provenance information, and license text along with the dataset to comply with terms.


\section{Statistical Analysis}
\label{sec:statistical_analysis}

To characterize the dataset, its inherent statistical utility, and correlation to not only guide development decision of compiler engineers, but also its utility for the training of large language models, a number of statistical analyses are performed. The ability to explore, and compare these analyses cross-language is a core novelty of our dataset to compiler engineers, as well as to the construction of machine-learned compiler componentry.

\subsection{Visualization of Properties}

The function properties are computed using the upstream
\mintinline{c}{FunctionPropertiesAnalysis} pass in LLVM, which we modified~\footnote{\href{https://reviews.llvm.org/D157358}{reviews.llvm.org/D157358}}~\footnote{\href{https://reviews.llvm.org/D158018}{reviews.llvm.org/D158018}}~\footnote{\href{https://reviews.llvm.org/D158681}{reviews.llvm.org/D158681}}
to give us a similar set of features to YaCoS~\citep{filho2018yacos}. To better understand the characteristics of the collected IR in terms of features of the underlying source code, and of the language itself, several analyses are performed.

\begin{figure}[H]
\centering
\includegraphics[width=\textwidth]{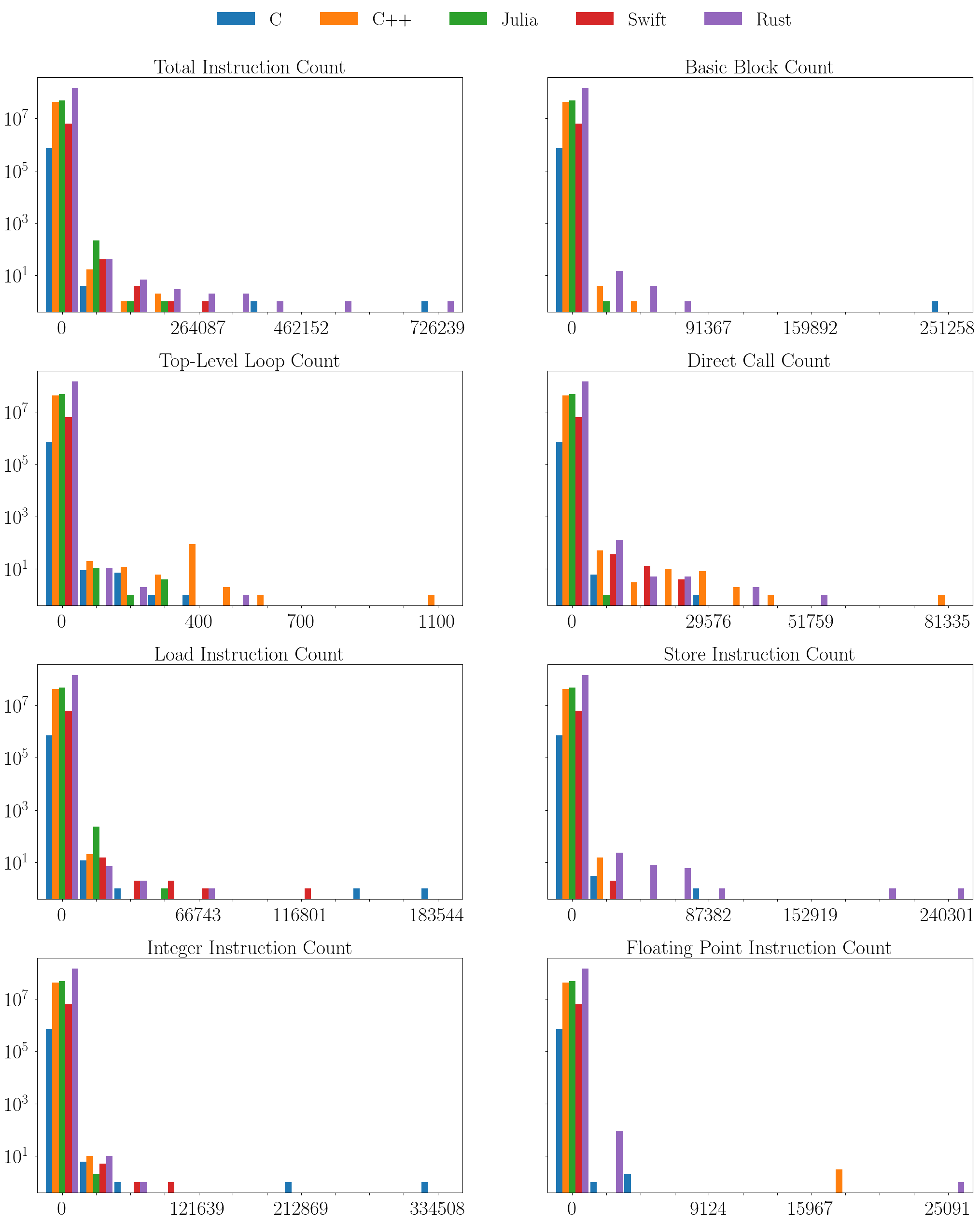}
\caption{
    Histograms of 8 different function properties. All function properties are analyzed across all 5 languages, and show a similar left-skew in their count-statistics.
}
\label{stats:properties_histograms}
\end{figure}

\subsection{Function Properties}

In addition to looking at the combination of function properties, we also looked at several function properties individually, comparing
them across languages as shown in \autoref{stats:properties_histograms}. We collected properties using LLVM's
\mintinline{c}{FunctionPropertiesAnalysis} pass, sampling 1,000,000 functions from each language contained within the dataset.
All of these variables show the same overall shape, a strongly
left-skewed distribution, but the exact characteristics are language dependent. In addition, most of these properties are correlated with
the length of the function under analysis, but show some distinct patterns depending upon the variable under analysis such as the load
instruction count and the floating point instruction count where certain languages have a significantly longer tail
than other languages, C and C++ for load instructions, and C for floating point instructions.
There are several other patterns such as the significantly longer tail for C++ in regards to direct calls, suggesting
small-function idioms. The long-tail of top-level loops in C/C++ also suggests some information about their usage.

\subsection{Function Duplication}

\begin{figure}[t]
\centering
\includegraphics[scale=0.7]{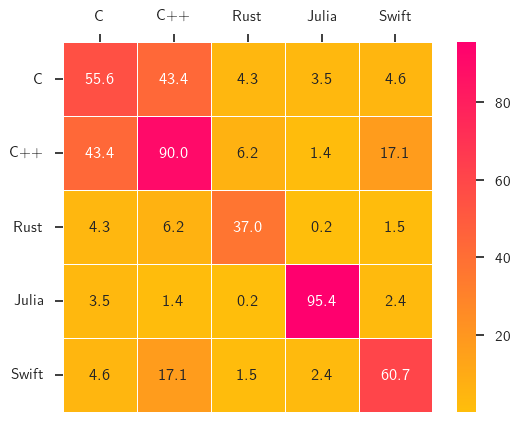}
\caption{
    Percentage of duplicate functions present between two languages as determined by the newly upstreamd
    \mintinline{c}{StructuralHash} in LLVM with detailed hashing enabled. All values are percentages.
}
\label{stats:duplication_heatmap}
\end{figure}

Furthermore, we performed an analysis quantifying function duplication within and between languages and present our results
in the heatmap shown in \autoref{stats:duplication_heatmap}. To compute function
duplication, we used a similar methodology to the one used for module-level deduplication in the initial deduplication stage.
We deduplicated using LLVM's \mintinline{c}{StructuralHash}, but for this analysis we looked at individual functions rather than the whole
module. Within the deduplicated data, some interesting patterns emerge. There is a much higher degree of duplication within
individual languages than there is between languages. We hypothesize this is caused primarily by the following two factors: language
idioms and function mangling. There are often a significant number of idioms within a language such as getters and setters in C++
that will often end up producing similar IR, causing a high degree of duplication within a language. In addition to this, different
languages use different mangling strategies, which significantly decreases the duplication rate between languages for functions
that involve function calls as \mintinline{c}{StructuralHash} takes function names into account when evaluating call instructions, 
on top of the names of the called functions potentially being different.
However, we do see more duplication between languages that share similar niches and compilation strategies. There is a significant
amount of overlap between C and C++ as they occupy similar software niches and, when compiled with clang, share a compiler frontend
and middle end. We also see some overlap between C++ and Swift, suggesting some similarity, potentially in language idioms. Next,
we observe that there is virtually no overlap between Julia and other languages, again supporting the hypothesis that Julia emits
code significantly different from other languages. Finally, we observe duplication between Rust and C++ in addition to a smaller
amount of duplication between Rust and C, but both are quite small.

\subsection{Opcode Distribution}

\begin{figure}[t]
\centering
\includegraphics[width=0.8\textwidth]{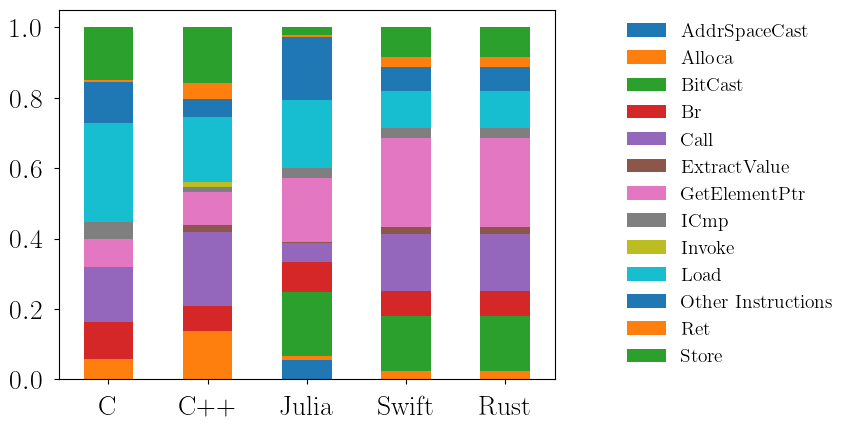}
\caption{
    LLVM IR opcode distribution of the top ten operations across all languages included in ComPile as computed by LLVM's \mintinline{c}{InstCount} pass.
}
\end{figure}

Next, we perform an analysis of the instruction distribution across languages. We use the LLVM \mintinline{c}{InstCount} pass
to count instructions at the module level and then aggregate the total number of instructions per language. This pass ignores
extraneous instructions like debug instructions, but does count some LLVM annotations presented as intrinsics, such as lifetime annotations,
as call instructions. There are several
interesting differences between the frontends for various languages that we observed. For example, we observed that Julia emits
significantly less store instructions than other languages, but takes significant advantage of instructions not within the ten
most frequent instructions compared to other languages. Other differences, such as the significant number of return instructions
in C++ suggest a large amount of small functions, which we also see to a lesser degree in Swift and Rust, which share some OOP idioms.
However, this could also be due to multi-exit functions as a choice of the program or as a pattern within the source code.
In addition, we see that certain languages are the only ones to use specific instructions. For example, Julia and Swift
make extensive use of the \mintinline{c}{BitCast} instruction while C, C++, and Rust do not use it. Finally, we observe an
increase in call instructions after optimization, which is most likely a result of inlining. It is important to note that an
increase in call instructions only describes the static count and the number of runtime call instructions would likely decrease.

\subsection{Token Count}

Finally, we performed experiments at different vocabulary sizes to gather approximate token counts to determine
the utility of our dataset for the code training of pre-trained large language models with the results shown in
\autoref{stats:public_tokens}. For these experiments we used the Llama 2 tokenizer~\citep{touvron2023llama} to be able to compare ComPile's size to contemporary datasets. To further test the size of the dataset, we generated
a vocabulary from a subset of the dataset. We chose to use BPE tokenization~\citep{sennrich2016subwordunits} as it is one of the most commonly used techniques
for tokenization for LLMs and easily adaptable to the textual component of our dataset.
We gathered approximately 400 bitcode modules from each language and disassembled
them into IR, training a BPE tokenizer over this data using fastBPE \footnote{https://github.com/glample/fastBPE},
generating several different vocabulary sizes for the various experiments. Finally, we used fastBPE to tokenize all the modules in our dataset after
disassembling them, counting the number of tokens generated and summing over the entire dataset. Relative to the large size of
our dataset in text form, approximately 2.8TB, we end up with comparatively few tokens. We believe this is primarily due to
the very formulaic nature of IR where there are many long character sequences that will occur often enough to be tokenized
into a single token. We note that this is a very naïve method of tokenizing a language as structured as LLVM-IR but believe
this serves as an appropriate estimate for the number of tokens that one could expect to obtain from our dataset.

\section{Related Work}
\label{sec:related_work}

There exist a number of related datasets of code for the training of machine learning models in literature.
Conceptually, we break these related datasets down into three main categories, as shown in \autoref{fig:related_work}. Case 1 consists
of datasets translating between two different codebases, case 2 considers reference work which translates
between two different languages by going through the IR as an intermediate translation step, case 3 consists
of a dataset of different languages without the structure to translate from one language, to the other explicitly,
and case 4 consists of a number of source languages, compiled to the IR, which, to our knowledge, only contains our dataset, ComPile,
and the dataset used in the work of~\cite{cummins2023llmcompileroptimizations}.
Most pretraining datasets for large language models~\citep{li2022competition,kocetkov2022stack,lozhkov2024starcoder2,markovtsev2018public}
fall into the third category, scraping source code from hosting services like GitHub, and GitLab with their expansive
index of individual repositories in all programming languages, and hence tend to produce very extensive datasets which
are only filtered for licensing issues, and then deduplicated, but do not take the quality of the included code into account.

\footnotetext[1]{This figure only includes the pretraining dataset for AlphaCode rather than the smaller competition sourced fine-tuning dataset.}
\footnotetext[2]{Size of training dataset not reported, and not reproducible. Dataset consists of 9.5B tokens, tokenized with fastBPE~\citep{sennrich2016subwordunits}.}

\begin{table*}[t]
    \centering
    \begin{tabular}{cccc}
    \toprule
    {\bf{Name of Dataset}} & {\renewcommand{\arraystretch}{1}\begin{tabular}[c]{@{}c@{}}\textbf{\textbf{Size}}\\\textit{(TB)}\vspace{.0em}\end{tabular}} & {\renewcommand{\arraystretch}{1}\begin{tabular}[c]{@{}c@{}}\textbf{\textbf{Programming}}\\\textbf{Languages}\vspace{.0em}\end{tabular}} & {\bf{Case}} \\
    \midrule
    The Stack & 2.9 & 358 Languages & Case 3 \\
    The Stack v2 & 32.1 & 358 Languages & Case 3 \\
    \rowcolor{col1}
    ComPile (closed) & 2.4 & Rust, Swift, Julia, C/C++ & Case 4 \\
    \rowcolor{col1}
    ComPile (public) & 1.9 & Rust, Swift, Julia, C/C++ & Case 4 \\
    Code Llama & 0.86 & $\leq$ 358 Languages & Case 3 \\
    TransCoder & 0.74 & C++, Java, Python & Case 1 \footnotemark[3] \\
    AlphaCode & 0.72 \footnotemark[1] & 12 Languages & Case 3 \\
    LLM for Compiler Opt. & 0.001 & C/C++ & Case 4 \\
    TransCoder-IR & \footnotemark[2] & C++, Go, Java, Rust & Case 2 \\
    HPCorpus & 0.07 & Fortran, C, C++ & Case 3 \\
    \bottomrule
    \end{tabular}
    \caption{
        Related datasets to our newly introduced dataset are the Stack~\citep{kocetkov2022stack}, the Stack v2~\citep{lozhkov2024starcoder2}, the datasets used for the training of Code Llama~\citep{roziere2023code}, TransCoder~\citep{lachaux2020unsupervised}, AlphaCode~\citep{li2022competition}, LLMs for Compiler Optimization~\citep{cummins2023llmcompileroptimizations}, and the HPCorpus~\citep{kadosh2023domain}. Code Llama, as well as AlphaCode, use filtered subsets of \href{https://console.cloud.google.com/marketplace/product/github/github-repos}{GitHub Activity Data}, where the filtering criteria of Code Llama are not known. We break all related datasets down into 4 distinct cases: \textbf{\emph{Case 1:}} The translation between two programming languages, \textbf{\emph{Case 2:}} The translation between two programming languages through the intermediate representation as an intermediate step, \textbf{\emph{Case 3:}} A mix of different codebases from different programming languages, and \textbf{\emph{Case 4:}} A mix of different codebases from different programming languages compiled to the intermediate representation. 
    }
    \label{fig:related_work}
    \vspace{-0.75cm}
\end{table*}

\noindent Relating to our dataset, datasets from the third class also do not guarantee that they, in themselves, are compilable, and
often contain auxiliary files such as documentation in Markdown. 
Another favored source of code within this category is
programming competitions, and while such code is inherently compilable, it bears little resemblance to code used in production.
In the case of modern large language models, the quality of the code is mitigated by only using the data for
fine-tuning~\citep{brown2020language}, or further instruction-tuning with reinforcement learning~\citep{ouyang2022training}
to achieve the desired downstream behaviour.
~\\

\noindent Case 1 contains a number of recent datasets for models which transcode between two programming languages,
examples of which include Transcoder~\citep{roziere2020unsupervised}, and recent efforts like the one of IBM
to translate COBOL to Java~\cite{ibm2023cobol}. Depending upon the specific methodology used for training,
datasets for this case can look similar to datasets for case 2 when techniques like back-translation are employed
for model training, but we make the distinction here primarily on dataset usage.
The extension of case 1 to translate between two different
programming languages by utilizing the intermediate representation as an intermediate translation step transforms
such a dataset into Case 2. The only example of this known to the authors is Transcoder-IR~\citep{roziere2023code}.

\noindent Complementary to these large pretraining-scale datasets, there exist a number of smaller, more focussed datasets aimed at the fine-tuning of already 
pretrained large language models~\citep{zhu2022xlcost,li2022competition,puri2021codenet}. These datasets are primarily collected through data extraction from 
coding competitions~\citep{li2022competition,puri2021codenet}, or the scraping of curated websites~\citep{zhu2022xlcost}. This guarantees a higher level of
quality in regards to buildability and structure for the included code, hence making them more optimal for fine-tuning. However, the data collection 
methodology implicitly introduces a lack of variety in the datasets. Coding competititon datasets might include a couple thousand coding exercises which 
contain a great many solutions to the same exercises, but yet they are only solving the very same set of coding problems. Optimizing for time-to-solution or
other narrow properties, such code also exhibits decidedly different characteristics to code used in production, hence making these datasets markedly 
different to ComPile.

\noindent Specifically for the task of machine-learned compiler heuristics, and machine-learned compiler componentry there exist
a number of statistics-focussed~\citep{kadosh2023quantifying}, compiler heuristics-focussed~\citep{armengol2022exebench},
and autotuning-focussed datasets~\citep{dasilva2021anghabench}. Often beginning with the web-scraping of large amounts of code,
these approaches modify the resulting code in a number of ways. Examples include the modification of arbitrary source files to make them 
compilable~\citep{dasilva2021anghabench}, executable~\citep{armengol2022exebench}, or abetting the statistical analysis of aspects of the 
code~\citep{kadosh2023quantifying}. ComPile, while being able to fulfill similar dataset demands, offers a number of key advantages.
The code in our dataset, by means of our dataset construction methodology, consists only of compilable code, using the same compilation
toolchain as used for production deployments, of which the IR is collected before optimization, allowing for IR at any stage of the
compilation pipeline to be easily generated. This allows ComPile to go significantly beyond the capabilities of previous
compiler-targeted datasets. 

\section{Limitations and Future Work}
\label{sec:limitations}

The presented dataset introduces a large corpus of compiled high-quality code. While this work has very
good coverage of languages such as Rust, Swift, and Julia, we had to make a number of implicit trade-offs
in the construction of our dataset. Compared to a number of other larger datasets such as the Stack 1 \& 2~\citep{kocetkov2022stack,lozhkov2024starcoder2},
we decided to not pursue a number of avenues to obtain the same order of magnitude of tokens, opening up a number of future avenues of work.

\noindent Following our approach to only include high-quality code in our dataset, we believe the
dataset could be significantly expanded by taking advantage of additional package ecosystems such as those of
Linux distributions. These ecosystems contain recipes with a consistent format across all packages that
describe the individual building steps and dependencies.
Some individual package ecosystems contain close to 1M recipes~\citep{AUR}, and could hence prove fruitful to the expansion of our IR database.
Adopting the widely used approach of GitHub repository scraping, we could also envision filtering the
list of repositories compilable with LLVM-IR generation, adding to the corpus only those repositories
of proven quality. Filtering parameters such as the number of contributors, number of commits, and other
activity metrics could remove outlier repositories that do not contain high-quality code (e.g., homework assignments and abandoned code).
However, we view this task as challenging due to these repositories not abiding by a consistent format,
having much higher variance in their build systems, and hence being much harder to compile with a single builder.
This problem extends to the dependencies of said GitHub repositories. Not all the projects have a consistent
dependency specification, hence requiring either manual dependency resolution or resolution in a highly complex automated fashion. 

\noindent Furthermore, we did not include a number of other language-specific ecosystems due to the inherent difficulties
and fragmentation of their build systems. Haskell, for example,  builds on the LLVM compilation infrastructure
and has a centralized package repository with specified dependencies. Nevertheless, including it in this dataset
proved infeasible due to the complexity of Haskell's dependency management which requires very specific Haskell
versions and highly specific dependencies with pre-specified versions. These convoluted dependencies are almost
impossible to be handled by the central builder schemes used in this project.
Other languages, like the HPC-language FORTRAN, did not see wider inclusion in the dataset due to the varied
compilation behavior of its multiple LLVM frontends. Added complexity came from highly specific compilation flags
for each FORTRAN project which vary from compiler frontend to compiler frontend and from build system to build system,
hence making it impossible to compile with one central compilation approach. While we could explore the integration of code from other datasets such as the Stack 2~\citep{lozhkov2024starcoder2},
coding competition datasets~\citep{li2022competition}, and HPC-focussed datasets of code~\citep{kadosh2023quantifying},
these would still be subject to the same restrictions as in the preceding paragraph. They would have to be filtered for
code quality, their build systems inspected for the amenability to a centralized builder, and for compiled languages 
such as C/C++ and FORTRAN their compilation would have to utilize a LLVM-based compiler.
~\\

\noindent In future work, we seek to expand upon our approach to more closely align large language models for code with compilation infrastructure. 
We will explore code-centric tokenization as an opportunity to closely incorporate knowledge about the programming language and IR structure.
We believe that by departing from textual tokenization of the IR, as is beginning to be explored in the
literature~\citep{guo2023lowlevelprograms,szafraniec2023transcoderir}, we can provide improved performance over the current state-of-the-art.
Our primary motivations for this belief include the fact that the distance between certain elements of IR, like attribute groups, and relevant
context, like their associated definitions, can easily exceed the context length of most LLMs, in addition to the possibility for more compact
tokenizations allowing more context to be given to models.
In using domain-specific tokenizations, we hope to preserve more of the semantic structure of the IR throughout the tokenization,
and hence improve downstream model performance. This 
approach will hopefully be able to yield small, performant models, which also retain the performance of larger models trained on textual tokenization.

\noindent In follow-up work, we seek to further explore the statistical properties of the dataset such as the distribution of code
within the dataset, the impact on the distribution of code by different collection methods, and the performance impact of
trained models these different distributions end up having. Further influences to be quantified are the influence of the
dataset construction techniques, the influence of the sources of the dataset on the distribution of the dataset, and most 
importantly the impact of distribution shifts in code datasets on downstream model performance.

\noindent With a fast moving project like LLVM, there are often significant changes between versions. Even within the rendition of the
dataset presented here, there is bitcode from multiple versions of LLVM depending upon the specific frontend used to compile
a piece of code. Many of these updates to LLVM involve significant changes to how IR should be produced by a compiler frontend,
such as the change from typed to opaque pointers. Having a static dataset means that the dataset becomes less relevant over time
as LLVM evolves further. We publish our tooling to produce the dataset which allows for the creation of a dataset similar to this
one produced with arbitrary frontend versions and leave it to future work to quantify the impact that distribution shifts over
time might have on the utility of a dataset such as this one.

\noindent Beyond the training of large language models, we also envision extensive future use of the dataset for the training of machine learned compiler components.
First, machine learned compiler heuristics have shown great promise~\citep{trofin2021mlgo}, but are held back by limitations of current datasets. ComPile
enables the metrics to be better trained, which has shown ample performance improvement in practice with performance metrics sometimes doubling from
having access to large amounts of previously hard to gather IR. Going beyond the improvement of existing machine learned compiler heuristics, the
presented dataset could furthermore make the training of heuristics such as e.g. the inlining-for-size in generic cases much easier. While learned
compiler heuristics only touch individual stages of the compilation pipeline, ComPile enables much more far-reaching work on
performance evaluations of LLMs as models of entire compilation pipelines, which has only been explored on small datasets previously~\citep{guo2023lowlevelprograms,mannarswamy2022mlinstcombine}.
ComPile is much broader in scope than the previously tested datasets, and our data collection approach allows for the collection of IR
at \textit{any} point of the compilation pipeline through simple postprocessing pipelines, hence enabling entirely new avenues of LLM 
compilation pipelines research.

\noindent The developed dataset collection, and compilation tooling is also to be further explored in future work. Its extensible pipeline
could potentially be used to automatically execute a plethora of unit tests and benchmarks throughout the build system, and hence
much better verify the individual stages of compilation and IR transformations. Extended with function instrumentation, and
replay tooling~\citep{casto2015cere}, future derivatives of ComPile could also include function inputs, and expected outputs along with
extracted functions to allow for fine-grained performance introspection on a grand scale, which is currently impossible. 
Potential future results of such fine-grained introspection throughout the compilation pipeline, and across programming
languages include better performance prediction without needing to compile a single line of a program, better evaluation of
the performance impact on individual compiler optimizations, and performance improvement through better compiler-generated code.

\section{Conclusion}
\label{sec:conclusion}
In this paper we present ComPile, a novel dataset of LLVM-IR collected from a number of package ecosystems
consisting of large production-grade codebases. It is significantly larger than previous finetuning-focussed, and compiler-focussed
code datasets, albeit smaller than large language model-focussed code pretraining datasets. Statistical analysis of the collected
dataset is performed, and differences in the IR properties of the collected IRs between languages are being shown.
ComPile's increased size in combination with its quality-focused construction methodology not only enables the systematic evaluation of
previous work, but opens up entirely new avenues of research for IR-centric machine learning, and most specifically machine-learned compiler
componentry for which the scale of this dataset paves the way to an entirely new generation of machine learning models for compilers.


\impact{%
The ComPile dataset will have the largest impact in machine learning for compilers, where it constitutes the first large language model scale dataset of compiler representation. It will hopefully enable wider progress in the application of machine learning to compilers, and the design thereof.
~\\

\noindent The potential uses of the dataset may benefit a large user base due the ubiquitous use of LLVM across compilers in industry such as the ones from Apple, Intel, IBM, and AMD, as well as LLVM's just-in-time (JIT) compiler, which used by a number of programming languages such as Julia, and the widely-used Python. Each optimization, heuristic, or learned pass ordering will in some capacity apply to all of these, and hence be for the broad benefit of all. While the, repeated, building of ComPile snapshots leads to a short-term increase in greenhouse gas emissions, the exact amount is hard to quantify due to varying datacenter efficiency. This is expected to be offset by the long-term benefits brought about by better compiler heuristics, and machine learning-improved compiler infrastructure trained on ComPile, whose impact is compounded by the widespread use of the LLVM compiler infrastructure.
~\\

\noindent There are important considerations made in the construction of ComPile to respect the licenses of the software packages ComPile is built from. The dataset is filtered for permissive licenses, as outlined in subsection~\ref{subsec:license_filtering}, and licenses are distributed alongside with the dataset. In addition, the public release of the dataset went through the rigorous internal release review of Lawrence Livermore National Laboratory (LLNL).
~\\

\noindent We would encourage further work into the biases inherent to the dataset, and its internal distribution of intermediate representation sources. Its construction is a conjunction of our best effort to represent the wider usage of LLVM across programming languages, and the ability to extract intermediate representation from centralized package indices. As such it is not representative of the wider usage of LLVM as outlined in section~\ref{sec:limitations}. To ensure the long-term benefit of ComPile, it's representative evaluation of the usage across languages utilizing the LLVM compiler infrastructure is going to be of the utmost importance. The exact impact of this is an open research question.
}


\acks{%
    \noindent We would like to thank Valentin Churavy for his assistance in understanding the Julia-compiler, and for volunteering his
    in-depth knowledge on pathways to the extraction of IR from Julia packages. We would also like to thank Todd Gamblin, Alec Scott, Harmen Stoppels,
    and Massimiliano Culpo for their assistance with Spack, and their prompt reviewing of our changes to Spack. We would furthermore like
    to express our gratitude to Nikita Popov, Arthur Eubanks, and all other LLVM contributors who helped with the reviews of the required
    patches to upstream LLVM.
    ~\\
    
    \noindent The views and opinions of the authors do not necessarily reflect those of the U.S. government or Lawrence Livermore National Security, 
    LLC neither of whom nor any of their employees make any endorsements, express or implied warranties or representations or assume any 
    legal liability or responsibility for the accuracy, completeness, or usefulness of the information contained herein.
    This work was in parts prepared by Lawrence Livermore National Laboratory under Contract DE-AC52-07NA27344 (LLNL-JRNL-854809).
}

\vskip 0.2in
\bibliography{references}

\appendix

\section{Dataset Statistics of the Closed Dataset}
\label{appendix:closet_dataset_stats}

\begin{table}[H]
    \centering
    \begin{tabular}{ccccccc}
        \multicolumn{7}{c}{\bf{\large{Closed Dataset}}} \\
        \toprule
        & \multicolumn{2}{c}{} & \multicolumn{4}{c}{\bf{BPE Tokens}} \\
        \cmidrule(l{5pt}r{5pt}){4-7}
         \multicolumn{2}{c}{\bf{Language}} & {\renewcommand{\arraystretch}{1}\begin{tabular}[c]{@{}c@{}}\textbf{\textbf{Llama 2 Tokens}}\\\textit{(billions)}\vspace{.0em}\end{tabular}} & {\renewcommand{\arraystretch}{1}\begin{tabular}[c]{@{}c@{}}\textbf{\textbf{10k}}\\\textit{(billions)}\vspace{.0em}\end{tabular}} & {\renewcommand{\arraystretch}{1}\begin{tabular}[c]{@{}c@{}}\textbf{\textbf{50k}}\\\textit{(billions)}\vspace{.0em}\end{tabular}} & {\renewcommand{\arraystretch}{1}\begin{tabular}[c]{@{}c@{}}\textbf{\textbf{50k}}\\\textit{(billions)}\vspace{.0em}\end{tabular}} & {\renewcommand{\arraystretch}{1}\begin{tabular}[c]{@{}c@{}}\textbf{\textbf{100k}}\\\textit{(billions)}\vspace{.0em}\end{tabular}} \\
         \midrule
         \cLogo & C & 16 & 3 & 2 & 2 & 1 \\
         \cppLogo & C++ & 116 & 30 & 17 & 14 & 12 \\
         \juliaLogo & Julia & 615 & 48 & 27 & 20 & 14 \\
         \rustLogo & Rust & 1079 & 198 & 132 & 116 & 102 \\
         \swiftLogo & Swift & 21 & 4 & 2 & 1 & 1 \\
         \midrule
         \multicolumn{2}{c}{\textbf{Total}} & \textbf{1848} & \textbf{282} & \textbf{179} & \textbf{154} & \textbf{130} \\
         \bottomrule
    \end{tabular}
    \caption{
        Token count of the encoded ComPile under varying vocabulary sizes, and considering the tokenization of the data with Byte-Pair encodings \emph{(BPE)}, and tokenization with the Llama 2 tokenizer.
    }
    \label{stats:closed_tokens}
\end{table}

\end{document}